\documentclass[prb,onecolumn,a4paper]{revtex4-1}

\usepackage{graphicx,setspace}
\usepackage{caption}
\large
\usepackage{amsmath}
\usepackage{float}
\usepackage{upgreek}
\usepackage{color}
\usepackage{soul}

\usepackage{epstopdf}
\usepackage{gensymb}
\usepackage[colorlinks = true,
			linkcolor = blue,
			urlcolor = blue,
			citecolor = blue,
			anchorcolor = blue]{hyperref}

\begin{document}


\title{Strong Exciton-Photon Coupling in Large Area MoSe$_2$ and WSe$_2$ Heterostructures Fabricated from Two-Dimensional Materials Grown by Chemical Vapor Deposition}

\author{Daniel J. Gillard$^{1\dagger}$}
\author{Armando Genco$^{1\ast\dagger}$}
\author{Seongjoon Ahn$^{2\dagger}$}
\author{Thomas P. Lyons$^1$}
\author{Kyung Yeol Ma$^2$}
\author{A-Rang Jang$^2$}
\author{Toby Severs Millard$^1$}
\author{Aur\'{e}lien A. P. Trichet$^3$}
\author{Rahul Jayaprakash$^1$}
\author{Kyriacos Georgiou$^1$}
\author{David G. Lidzey$^1$}
\author{Jason M. Smith$^3$}
\author{Hyeon Suk Shin$^2$}
\author{Alexander I. Tartakovskii$^{1\ast\ast}$}

\medskip

\affiliation{$^1$Department of Physics and Astronomy, University of Sheffield, Sheffield S3 7RH, UK}

\smallskip

\affiliation{$^2$Department of Energy Engineering, Department of Chemistry, and Low Dimensional Carbon and 2D Materials Center, Ulsan National Institute of Science and Technology, Ulsan 44919, South Korea}

\smallskip

\affiliation{$^3$Department of Materials, University of Oxford, Parks Road, Oxford, OX1 3PH, UK}

\smallskip

\affiliation{$^{\dagger}$These authors contributed equally to this work.}

\smallskip

\author{E-mail addresses: dgillard1@sheffield.ac.uk}
\author{$^\ast$a.genco@sheffield.ac.uk}
\author{sjahn@unist.ac.kr}
\author{t.lyons@sheffield.ac.uk}
\author{kyma@unist.ac.kr}
\author{arjang@unist.ac.kr}
\author{t.seversmillard@sheffield.ac.uk}
\author{aurelien.trichet@materials.ox.ac.uk}
\author{r.jayaprakash@sheffield.ac.uk}
\author{k.georgiou@sheffield.ac.uk}
\author{d.g.lidzey@sheffield.ac.uk}
\author{jason.smith@materials.ox.ac.uk}
\author{shin@unist.ac.kr}
\author{$^{\ast\ast}$a.tartakovskii@sheffield.ac.uk}

\affiliation{Corresponding authors: $^\ast$a.genco@sheffield.ac.uk, $^{\ast\ast}$a.tartakovskii@sheffield.ac.uk}

\date{\today}
\maketitle
\pagebreak

\section*{Abstract}
\textbf{Two-dimensional semiconducting transition metal dichalcogenides embedded in optical microcavities in the strong exciton-photon coupling regime may lead to promising applications in spin and valley addressable polaritonic logic gates and circuits. One significant obstacle for their realization is the inherent lack of scalability associated with the mechanical exfoliation commonly used for fabrication of two-dimensional materials and their heterostructures. Chemical vapor deposition offers an alternative scalable fabrication method for both monolayer semiconductors and other two-dimensional materials, such as hexagonal boron nitride. Observation of the strong light-matter coupling in chemical vapor grown transition metal dichalcogenides has been demonstrated so far in a handful of experiments with monolayer molybdenum disulfide and tungsten disulfide. Here we instead  demonstrate the strong exciton-photon coupling in microcavities comprising large area transition metal dichalcogenide / hexagonal boron nitride heterostructures made from chemical vapor deposition grown molybdenum diselenide and tungsten diselenide encapsulated on one or both sides in continuous few-layer boron nitride films also grown by chemical vapor deposition. These transition metal dichalcogenide / hexagonal boron nitride heterostructures show high optical quality comparable with mechanically exfoliated samples, allowing operation in the strong  coupling regime in a wide range of temperatures down to 4 Kelvin in tunable and monolithic microcavities, and demonstrating the possibility to successfully develop large area transition metal dichalcogenide based polariton devices.}

\section*{Introduction}

Monolayers of transition metal dichalcogenides (TMDs) are promising semiconductors with unique electrical and optical properties arising from the quantum confinement experienced by the electrons and holes in the two-dimensional (2D) structure\cite{Novoselov2005,Mak2016}. One of the main effects is the appearance of direct bandgap excitonic transitions showing features strongly beneficial for optoelectronics, such as high binding energies and very large oscillator strengths\cite{Mak2013,He2014}. Moreover, the breaking of spatial inversion symmetry in the 2D lattice and a large spin-orbit coupling generate spin-valley locked optically addressable excitons at the K and K’ points of the momentum space \cite{Mak2013,Xu2014}. These exceptional properties can be further enriched by integrating the TMDs within optical resonators enabling the strong exciton-photon coupling regime, where confined photons and excitons hybridize into new states called polaritons \cite{Liu2015,Dhara2018,Dufferwiel2015,Sidler2017,Low2016,Schneider2018,Krol2020}. Polaritons in TMDs acquire novel properties arising from the valley pseudo-spin degree of freedom of excitons \cite{Dufferwiel2017,Chen2017}, and further provide enhanced valley coherence for excitons strongly coupled with long-lived cavity photons \cite{Dufferwiel2018,Qiu2019}. Efficient polariton propagation in TMDs has been recently observed, with diffusion lengths of up to 20 $\mu$m in WS$_2$ at room temperature \cite{barachati2018interacting}, while valley-dependent divergent polariton diffusion has been found in MoSe$_2$ at cryogenic temperatures, whereby polaritons spread in different in-plane directions owing to the exciton valley Hall effect \cite{lundt2019optical}. Polaritons in TMDs already offer the potential to create highly non-linear phenomena \cite{Waldherr2018,Emmanuele2020,Dhara2018}, with further effort directed at realization of Bose-Einstein condensation\cite{Kasprzak2006,Waldherr2018}, polariton lasing \cite{Christopoulos2007,Bhattacharya2014} and optical parametric oscillation \cite{Amo2009} so far observed in other material systems. United to the valley degree of freedom of TMD monolayers, these phenomena could be exploited to create large scale all-optical polariton circuits and quantum networks\cite{Liew2010,Ballarini2013}.

However, TMD monolayers are usually fabricated by mechanical exfoliation, resulting in high quality but small sized flakes, hindering the reproducibility and scalability of the devices. Chemical vapor deposition (CVD) offers an alternative growth and fabrication method that provides substrate-wide coverage of uniform monolayer islands\cite{Shree2019}, as well as the ability to grow heterostructures in-situ\cite{Zhang2019a}, therefore completely bypassing the mechanical transfers necessary in exfoliated equivalents and minimizing external contamination. CVD provides a very attractive and scalable method for the fabrication of large scale TMD based devices. As such, CVD-grown MoS$_2$ and WS$_2$ flakes have already been employed in polaritonic devices working at room temperature\cite{Liu2015,Gebhardt2018}. Nevertheless, in order to optimize the polariton valley properties and optimize coherence, narrow exciton linewidths (low inhomogeneous broadening) and low structural disorder are needed, which up to now have only been shown by exfoliated MoSe$_2$ and WSe$_2$ monolayers at low temperatures \cite{Dufferwiel2015,Dufferwiel2017,Dufferwiel2018,Sidler2017,Low2016,Schneider2018,Dhara2018}. Another advantage of high quality structures and the possibility to operate at low temperature will be the access to highly non-linear trion-polaritons \cite{Emmanuele2020,Dhara2018} or Fermi-polaron-polaritons \cite{Sidler2017}, requiring control and stability of the charged excitons, as well as employment of suitable heterostructures \cite{Sidler2017} comprising of TMDs, hexagonal boron nitride (hBN) and graphene.  

Here, we show large area MoSe$_2$ and WSe$_2$ monolayers encapsulated in hBN, all grown by CVD, with crystal domains exceeding 100 $\mu$m in size, which display high optical quality, rivalling exfoliated material. This is evidenced by the intense and narrow exciton peaks observed in photoluminescence (PL) measurements and in reflectance contrast (RC). Substrate-wide growth uniformity and a high degree of alignment within the ensemble of monolayer domains has been achieved in these materials using CVD growth on sapphire for MoSe$_2$ and directly on CVD-synthesized hBN for WSe$_2$. We have confirmed these favorable monolayer TMD crystal properties by using a recently developed substrate-wide statistical analysis of TMD crystal axes orientation and PL properties \cite{SeversMillard2020}. Further testing of the samples in a tunable microcavity exposes clear evidence of strong exciton-photon coupling and formation of exciton-polariton states. An anti-crossing is observed between the cavity mode and neutral exciton transition, with Rabi splittings of about 17 meV for both MoSe$_2$ and WSe$_2$, very similar in magnitude with exfoliated materials \cite{Dufferwiel2015,Sidler2017,Dufferwiel2017,Dufferwiel2018}, highlighting similar optical qualities. Finally, as a proof of concept for future large scale TMD-based polaritonic devices, a monolithic cavity is fabricated incorporating CVD grown hBN-encapsulated MoSe$_2$ in a SiO$_2$ spacer, sandwiched between two mirrors, in which strong coupling is observed with a large Rabi splitting of 34 meV at T=4 K and 31 meV at T=150 K. The results presented in this work demonstrate the possibility to fabricate large area polariton devices exploiting high quality TMD based heterostructures made from CVD-grown materials, paving the way for future scalable TMD-polaritonic circuits.

\section*{Results}

\begin{figure}[t!]
	\centering
	\includegraphics[width=0.95\textwidth]{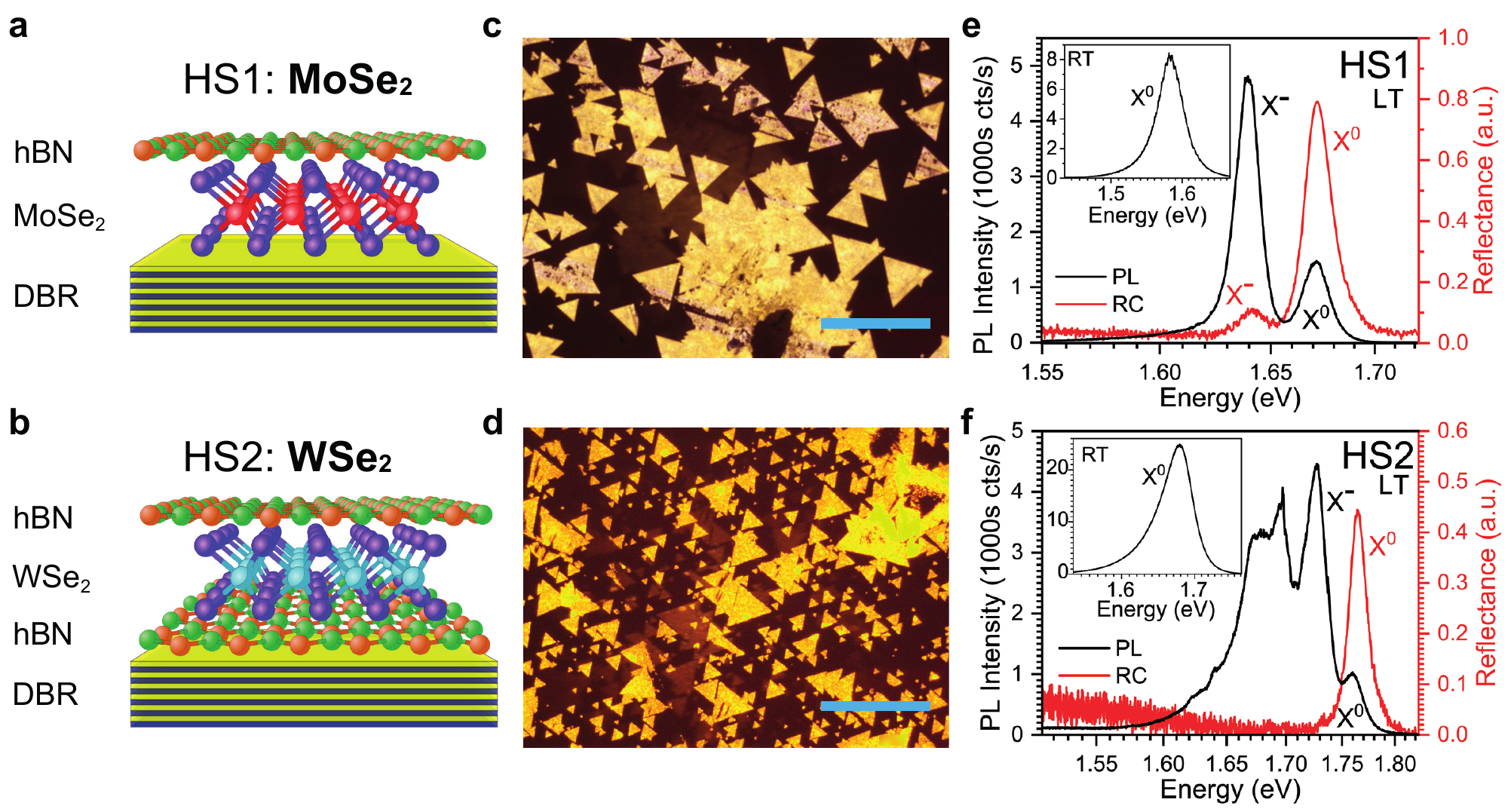}
	\caption{\label{Fig 1}{\bf a, b)} Schematics of the CVD grown heterostructures (not to scale). a) HS1: MoSe$_2$ monolayers were grown by CVD on a sapphire substrate, mechanically transferred onto a DBR mirror and encapsulated with monolayer hBN. b) HS2: WSe$_2$ monolayers were grown by CVD on multilayer hBN, transferred together with the hBN onto a DBR mirror and encapsulated with multilayer hBN. The hBN multilayers were also grown by CVD. {\bf c, d)} Optical microscope images of the photoluminescence (PL) from c) HS1 and d) HS2 taken at room temperature using a 50x objective lens (scale bar: 50 $\mu$m). {\bf e, f)} Optical characterization of e) HS1 and f) HS2 at T$\sim$ 4 K. Insets show PL spectra at room temperature. PL and reflectance contrast (RC) spectra are shown in black and red, respectively. At low T, both samples show emission from both neutral, $X^0$, and charged, $X^-$, excitons, whereas only HS1 shows absorption from both. PL cw excitation conditions: $\lambda_{exc}$ = 660 nm, P = 20 $\mu$W for HS1; $\lambda_{exc}$ = 532 nm, P = 20 $\mu$W for HS2.}
\end{figure}

Two types of encapsulated heterostructure (HS) have been fabricated from CVD-grown materials for this work (see further details in Methods). In the first heterostructure, HS1 (Fig.\ref{Fig 1}(a)), MoSe$_2$ monolayers are grown by CVD onto a sapphire substrate. Thousands of individual monolayer islands, positioned throughout the entire substrate, are then mechanically transferred using a polystyrene membrane onto a high reflectivity distributed Bragg reflector (DBR) composed of 13 pairs of SiO$_2$/Ta$_2$O$_5$ with the high reflectance stop-band centered at a wavelength of 750 nm. A substrate-wide few-layer film of hexagonal boron nitride (hBN), also grown by CVD, is mechanically transferred on top of the MoSe$_2$ flakes to complete the encapsulation. For the second heterostructure, HS2 (Fig.\ref{Fig 1}(b)), WSe$_2$ monolayers are grown by CVD directly on few-layer hBN, also grown by CVD. Both materials, WSe$_2$/hBN, are then mechanically transferred, at once, onto the DBR and subsequently encapsulated with a few-layer film of CVD grown hBN. It has been shown that for mechanically exfoliated TMDs, encapsulating with thin hBN provides uniform dielectric screening of the Coulomb interaction, reducing spatial inhomogeneity in the exciton, thereby narrowing the emission linewidth \cite{Ajayi2017,Shree2019}. Furthermore, hBN protects the TMD layers in the heterostructure from damage and contamination during the subsequent deposition of various dielectrics in microcavity structures relevant to our work \cite{DelPozoZamudio2020}. Moreover, the direct growth of TMD monolayers on hBN is also strongly beneficial as a route to single-crystal epitaxial growth \cite{Zhang2019,SeversMillard2020}, which, up until now, has been demonstrated with a limited range of TMD materials such as WS$_2$\cite{Okada2014} and MoS$_2$ \cite{Yu2017a}. Fully coalesced CVD grown WSe$_2$ monolayer films on hBN were obtained very recently by Zhang et al \cite{Zhang2019}, through a careful control of nucleation and extended lateral growth time, and a strong improvement of optical and electrical properties have been achieved compared to the same material grown on sapphire.

As a first characterization step, the room temperature PL emission and the general morphology of the structures have been analyzed under an optical microscope (see Methods). HS1 (Fig.\ref{Fig 1}(c)) generally consists of large isolated monolayer islands with characteristic triangular shape and average lateral size of 8 $\mu$m (see Supplementary Note I for details), along with a number of regions where multiple flakes merge to form monolayers with sizes exceeding 100 $\mu$m. Similarly, HS2 (Fig.\ref{Fig 1}(d)) shows large triangular monolayer flakes with average lateral widths of 11 $\mu$m. Again, in areas where flakes merge, monolayers of over 100 $\mu$m in width can be seen. Large areas of uniform coverage are necessary for constructing large arrays of identical heterostructure devices, such as transistors or photodetectors. Overall, there is monolayer coverage of 14\% for HS1 and 22\% for HS2, calculated by dividing the total area of monolayer across the sample by the total substrate area. The substrate-scale PL imaging analysis \cite{SeversMillard2020} used to identify the monolayers is discussed in more detail below.

Further optical characterization of the two heterostructures has been performed using a spectroscopic microscopy setup at both room ($\sim$ 290 K) and low ($\sim$ 4 K) temperature (Figs.\ref{Fig 1}(e, f)). PL measurements are carried out by exciting the samples with an off resonant continuous wave (cw) laser at a power of 20 $\mu$W (see Methods). The considerable room temperature excitonic PL emission shown in the insets in Figs.\ref{Fig 1}(e, f) highlights the large exciton binding energy associated with TMD monolayers \cite{He2014}. In HS1 the exciton PL peak is located at 1.579 eV with a linewidth of 40 meV, and HS2 displays the exciton peak at 1.670 eV with a similar linewidth of 45 meV, typical of MoSe$_2$ and WSe$_2$ monolayers operating at room temperatures. 

Decreasing the temperature to $\sim$ 4 K produces a narrowed and blue shifted neutral exciton PL peak, $X^0$, at 1.671 eV in HS1 and 1.759 eV in HS2, and a second peak attributed to the charged exciton (trion) transition, $X^-$, which appears at 1.639 eV in HS1 and 1.726 eV in HS2, about 30 meV below the neutral exciton \cite{Ross2013,Dufferwiel2015,Dufferwiel2017}. The relative intensity of the $X^-$ peak, when compared to the $X^0$ peak, is heavily influenced by the free carrier densities present in the structures\cite{Mak2013}. In HS2, PL seen at lower energies (below 1.70 eV) has previously been attributed to various excitonic complexes in WSe$_2$ including spin dark excitons \cite{Robert2017}, exciton-phonon side-bands\cite{Lindlau2018} and localized states\cite{Jadczak2017}. The samples show neutral exciton linewidths of 13 meV and 21 meV for HS1 and HS2 respectively, and 12 meV and 20 meV for the charged exciton of the two samples. Generally, the linewidth of an excitonic transition in TMD monolayers is strongly affected by the level of structural disorder and density of defects \cite{Gebhardt2018}. The spectral shapes and linewidths demonstrated by the CVD grown samples investigated in this work improve upon those reported by Zhang et al\cite{Zhang2019} and Lippert et al\cite{Lippert2017}, and are similar to exfoliated flakes operating at low temperature without encapsulation \cite{Ross2013,Ajayi2017,Shree2019,Dufferwiel2018,Dufferwiel2015}. This shows that CVD growth can produce heterostructures of comparable optical quality to mechanically exfoliated flakes. The role of hBN is mostly to provide a high quality substrate for the TMD synthesis\cite{Zhang2019,SeversMillard2020}, but also to act as a buffer layer protecting TMDs from damage during the deposition of additional layers in order to complete a microcavity or waveguide structure.

We also measure reflectance contrast spectra using a broad band white light source and calculated as $\Delta$ $R/R= (R_{sub}-R_{HS}) / R_{sub}$, where $R_{HS}$ is the reflectance of the heterostructure, and $R_{sub}$ is the reflectance of the bare substrate. These spectra (red lines in Figs.\ref{Fig 1}(e, f)) reveal a strong absorption peak attributed to the $X^0$ transition in both heterostructures and a lower intensity peak at lower energy attributed to $X^-$ in HS1. The relative peak height is strictly related to the oscillator strength of individual  transitions, with the neutral exciton being much more intense than the trion in HS1 due to a relatively low doping level.

\begin{figure}[t!]
	\centering
	\includegraphics[width=0.95\textwidth]{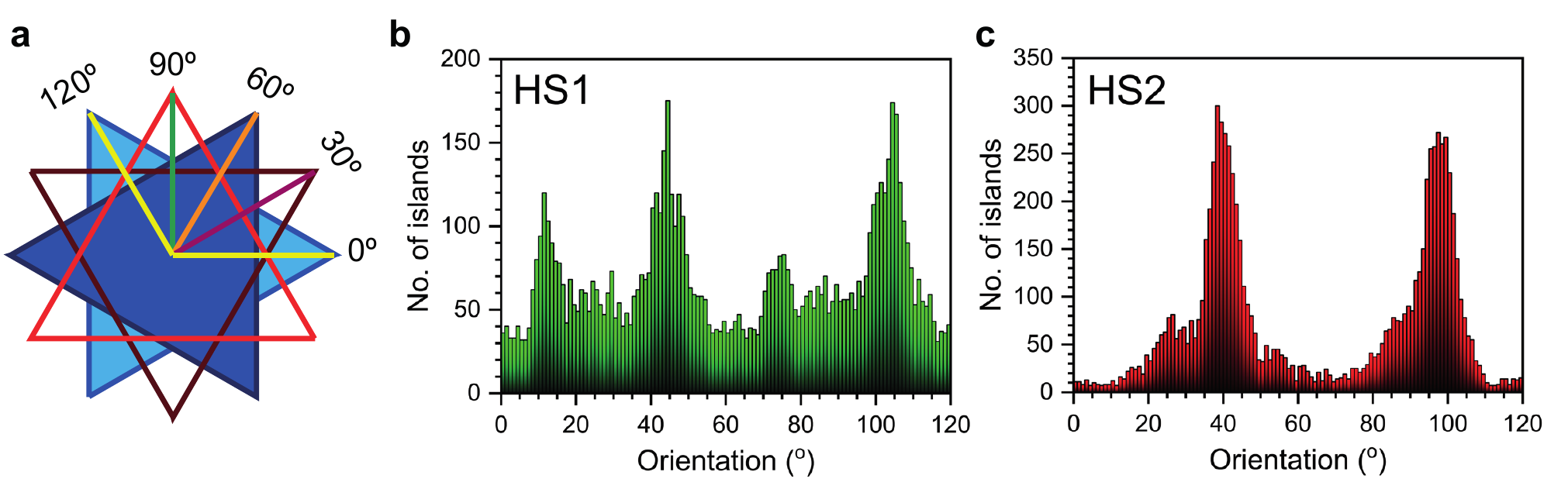} 
	\caption{\label{Fig 2} {\bf a)} Schematic showing flake orientations at different angles relative to the horizontal axis of the microscope images. {\bf b, c)} Analysis of flake orientation. Islands of monolayer TMD are identified and the orientation extracted using methodology as described in previous investigations\cite{SeversMillard2020}. Both b) HS1 and c) HS2 show two main peaks situated 60$\degree$ apart, equivalent to two opposite growth configurations rotated by 180$\degree$ due to the three fold symmetry of equilateral triangles. HS1 also shows two extra peaks situated at 30$\degree$ from the main peaks.}
\end{figure}

As can be inferred in Figs.\ref{Fig 1}(c, d), the bright triangular monolayer islands appear to have a preferred orientation. To extract the size and shape of monolayer flakes, shape recognition techniques were used on a full substrate map comprised of multiple microscope PL images, an example of one such image is shown in Figs.\ref{Fig 1}(c, d). By employing analytical methods detailed in our previous work \cite{SeversMillard2020}, the flake orientation relative to the horizontal axis of the microscope images can be found (Fig.\ref{Fig 2}(a)). In order to maximize accuracy, only islands with shape close to equilateral triangles are analyzed in terms of angular orientation. Of the 16999 (14205) individual monolayer islands identified in HS1 (HS2), 8089 (8391) satisfy this condition. Measured in terms of area, this corresponds to 21\% (17\%) of the total monolayer coverage. The histograms in Figs.\ref{Fig 2}(b, c) detail the number of islands identified as a function of orientation angle, showing that both the samples feature a very high degree of island orientation uniformity, a signature of epitaxial growth. For WSe$_2$ grown directly onto hBN (Fig.\ref{Fig 2}(c)), two main orientations have been found. This is expected from a sample with a three-fold symmetric triangular morphology presenting two possible opposite growth directions at 180$\degree$ to one another. These two preferential directions are directly related to the hexagonal crystal structure of the growth substrate and have also been observed in previous studies of MoS$_2$, WS$_2$, and WSe$_2$ grown by CVD on hBN\cite{Okada2014,Yu2017a,Zhang2019,SeversMillard2020}. For MoSe$_2$ grown onto c-plane sapphire (Fig.\ref{Fig 2}(b)), four peaks in the angular distribution are observed. Two main peaks show the preferred flake orientation, situated at 60$\degree$ relative to each other, along with two less populated angles at 30$\degree$ relative to the two main peaks. Both two, and four preferential growth directions have been seen in TMDs grown via CVD onto c-plane sapphire\cite{Dumcenco2015,Zhang2018}. Control over the relative angle of the flakes at the synthesis stage of fabrication will provide the basis to build scalable heterostructures with control over the relative interlayer crystallographic orientation. 

\begin{figure}[t!]
	\centering
	\includegraphics[width=.95\textwidth]{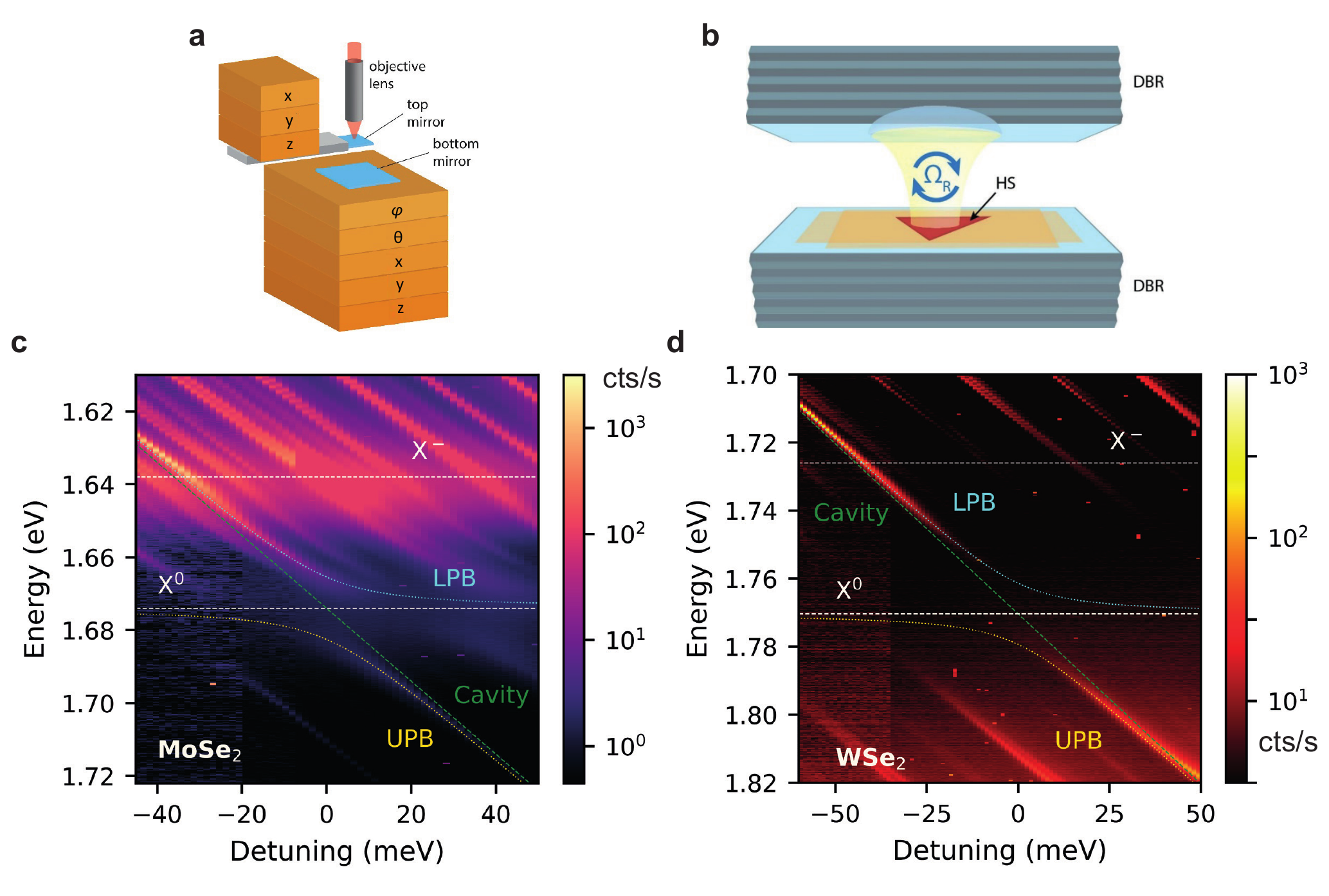}
	\caption{\label{Fig 3} {\bf a)} Schematic of the tunable open microcavity including the set of piezo actuators used to align the mirrors and perform the PL scans of the heterostructure. {\bf b)} Schematic of the open optical microcavity. The cavity is composed of a planar DBR, upon which the HS is placed, and a concave top DBR confining the optical cavity mode in 3 dimensions. {\bf c, d)} PL emission from c) HS1 and d) HS2 displayed as a function of photon energy and exciton-photon detuning ($\Delta = E_c - E_{X^0}$). Clear anti-crossings of the cavity mode with the exciton are observed in both heterostructures. PL spectra are fitted using a Lorentzian peak (see Supplementary Note II) and a two level coupled oscillator model is used to extract the lower (blue curve) and upper (yellow curve) polariton branches, excitonic resonances (white horizontal lines), and LG$_{00}$ photonic mode (green diagonal line). Rabi splittings of about 17 meV are found for both HS1 and HS2. Samples are optically excited using a 637 nm cw laser.}
\end{figure}

After the optical characterization step, the encapsulated heterostructures are tested in a tunable open cavity setup which consists of a top concave DBR mirror distanced 2-3 $\mu$m from a planar bottom DBR mirror (Fig.\ref{Fig 3}(a, b), upon which the HSs are placed \cite{Schwarz2014,Dufferwiel2015,Dufferwiel2017,Sidler2017,Dufferwiel2018}. The mirrors are positioned using piezo-actuator stages (Fig.\ref{Fig 3}(a)). Free space optical access from above the top concave DBR allows laser excitation and optical detection, using an achromatic doublet objective lens. A cavity length can be tuned by slowly moving the bottom mirror along the {\it z}-axis, thus allowing the tuning of the cavity mode (diagonal green dashed lines in Figs.\ref{Fig 3}(c, d)). The PL signal collected from the TMDs, as the cavity length is reduced, is displayed in Figs.\ref{Fig 3}(c, d)) as a function of detuning ($\Delta$, the energy difference between the cavity mode and unperturbed exciton, $X^0$). The three dimensional optical confinement provided by the concave top mirror generates a set of transverse modes for each longitudinal mode\cite{Schwarz2014,Dufferwiel2015} also visible in both figures. 

When the fundamental longitudinal cavity mode, LG$_{00}$ (green diagonal dashed lines in Figs.\ref{Fig 3}(c) and (d)) which ensures the highest light confinement, is tuned into resonance with the exciton transition energies, light matter coupling can manifest in one of two ways, both of which are observed in HS1 (Fig.\ref{Fig 3}(c)). As the cavity mode is tuned into resonance with the trion, $X^-$, at 1.638 eV, the mode is broadened and brightened, a demonstration of the weak coupling \cite{Dufferwiel2015,Dufferwiel2017,Dufferwiel2018,Kavokin2008} between the cavity and the trion transition with small oscillator strength. The absence of mode broadening in HS2 (Fig.\ref{Fig 3}(d)) cavity scans at the trion energy is an indication that the absorption of the WSe$_2$ trion resonance, occurring in HS2 at 1.726 eV, is too weak, as also confirmed in Fig.\ref{Fig 1}(f)).

The second regime of exciton-photon coupling, known as the strong coupling, presents itself as an anti-crossing of the LG$_{00}$ cavity mode and the exciton energies with a characteristic Rabi splitting, {\it $2\hbar\Omega_R$}, at the resonance. This behaviour can be clearly observed in both HS1 (Fig.\ref{Fig 3}(c)) and HS2 (Fig.\ref{Fig 3}(d)) as the LG$_{00}$ is tuned into resonance with the neutral exciton, at 1.673 eV for HS1 and 1.770 eV for HS2. The peak positions of the lower (LPB) and upper (UPB) polariton branches have been extracted using a Lorentzian peak fitting, and used to fit a two-level coupled oscillator model (detailed in Supplementary Note II) in order to determine {\it $2\hbar\Omega_R$} as shown overlaid in Figs.\ref{Fig 3}(c, d). We find a value of 17.2 $\pm$ 3.3 meV for HS1 and 16.8 $\pm$ 3.1 meV for HS2.

These measurements demonstrate large Rabi splittings closely comparable to the values observed in exfoliated monolayer MoSe$_2$ and WSe$_2$ placed in zero-dimensional tunable microcavities \cite{Schwarz2014,Dufferwiel2017,Sidler2017,Dufferwiel2018}. This further confirms, thanks to the reduced structural disorder in the presented heterostructures, the high optical quality of the hBN encapsulated CVD grown TMD monolayers, hence proving the validity for CVD growth techniques when designing scalable devices. 

\begin{figure}[t!]
	\centering
	\includegraphics[width=0.95\textwidth]{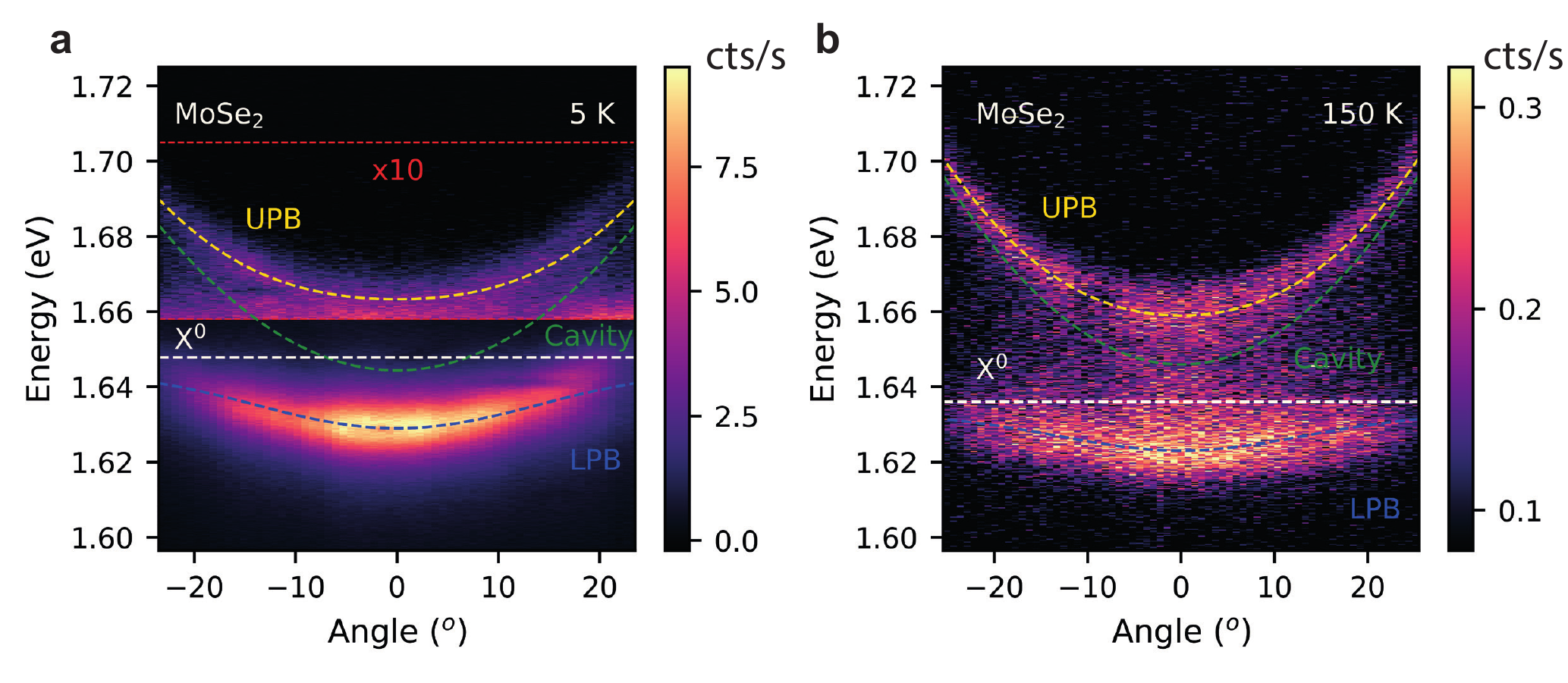}
	\caption{\label{Fig 4} {\bf a, b)} Angle resolved PL imaging of HS1 monolithic cavity. a) At 5 K, the cavity has a negative detuning at 0$\degree$  of $\Delta \approx$ -3.5 meV, showing anticrossings at $\pm$ 7$\degree$. The PL intensity in the region within the red dashed lines has been multiplied by a factor of 10 for clarity. b) At 150 K, the cavity has a positive detuning at 0$\degree$ of $\Delta \approx$ 10 meV. We overlay the fitted LBP (blue) and UPB (yellow) along with the extracted parabolic cavity dispersion (green) and neutral exciton resonance (white). A Rabi splitting of 34 $\pm$ 4 meV and 31 $\pm$ 4 meV is observed at 5 K and 150 K respectively. The sample is optically excited using a 660 nm cw laser at a power of 50 $\mu$W.}
\end{figure}

Further advantages of large area TMDs can be exploited in monolithic cavities, providing a platform to form various topological designs to adapt or enhance device functionality towards polariton circuits \cite{Liew2010,Ballarini2013}. For these devices, the protective function of the hBN encapsulation is of particular importance as the top dielectric mirror needs to be deposited on top of the TMD layers \cite{DelPozoZamudio2020}. As a proof of concept, we deposited 98 nm of SiO$_2$ via e-beam deposition, followed by a semi-transparent layer of 50 nm gold, on top of HS1 to fabricate a $\lambda$/2 monolithic cavity (see Supplementary Note III). The oxide deposition process has been carried out at room temperature in order to preserve the optical integrity of the emitting materials as much as possible \cite{GENCO2018174,DelPozoZamudio2020}. 

In the tunable cavity, the photonic modes are confined in all three dimensions, resulting in a set of discrete cavity modes, with k$_{x,y,z}$ $\sim$ 0, which are tuned in energy by altering the cavity length. By contrast, in a monolithic two-dimensional cavity, as in our case, the photonic mode is confined only in the vertical {\it z} direction, and thus a cavity mode energy dispersion as a function of continuous k$_{x,y}$ values is observed\cite{Kavokin2008}. This dispersion can be probed by measuring angle-resolved PL or reflectivity spectra as a function of angle measured from the normal to the sample (corresponding to k$_{x,y}$=0)\cite{Kavokin2008}. In the cavity used in our experiment, a stronger light confinement can be achieved than in the tunable devices presented earlier, due to a smaller thickness of the cavity spacer and lower mode penetration into the top mirror. As shown below, this leads to a higher magnitude of the Rabi splitting. Since monolithic cavities are not tunable in size, the temperature dependence of the $X^0$ transition energy (further discussed in Supplementary Note III) is used to tune the exciton into resonance with the photonic mode which has a negligible dependence of its frequency with temperature.

The PL collected from the monolithic cavity while being optically excited by non-resonant continuous wave laser, as imaged by angle resolved spectroscopy, is shown in Fig.\ref{Fig 4}. In Fig.\ref{Fig 4}(a) at a temperature of 5 K, the $X^0$ transition is at 1.648 eV, while in Fig.\ref{Fig 4}(b) at 150 K, the $X^0$ red-shifts to 1.636 eV. The monolithic cavity shows strong exciton-photon coupling signatures in PL at both the temperatures, owing to the protective capability of the CVD grown hBN which helped shield the MoSe$_2$ monolayers from the potentially damaging SiO$_2$ deposition process. 

At a temperature of 5 K, the exciton is negatively detuned from the cavity mode at 0$\degree$ by - 3.5 meV, such that the LPB is much more visible than the UPB. To show the upper polariton branch in Fig.\ref{Fig 4}(a) the collected intensity values have been multiplied by a factor of 10 between 1.658 eV and 1.705 eV, as outlined by red dashed lines. By fitting the PL emission spectra with Lorentzian peaks and applying the extracted peak positions to a two level coupled-oscillator model (see Supplementary Note II)\cite{Kavokin2008} we obtain a large Rabi splitting of 34 $\pm$ 4 meV at 5 K. The exciton-photon coupling strength in the monolithic cavity is higher than in the open cavity due to the increased light confinement. The two polariton branches, cavity dispersion, and exciton energy obtained from the coupled-oscillator model are shown overlaying Figs.\ref{Fig 4}(a) and \ref{Fig 4}(b), with an anti-crossing clearly seen at $\pm$ 7$\degree$ when the device is at 5 K. The strongly coupled cavity performs well up to 150 K, when the excitonic mode is positively detuned ($\Delta(0^o)$ = + 10 meV), leading to a Rabi splitting of 31 $\pm$ 5 meV.

\section*{Discussion}

In summary, high quality substrate-wide MoSe$_2$, and WSe$_2$, TMD monolayers encapsulated with large area hBN were fabricated using CVD growth techniques, and subsequently embedded in tunable and monolithic microcavity devices where strong exciton-photon coupling was observed. The heterostructures show optical properties comparable with exfoliated materials, and consequently exhibit similar values of polariton Rabi splittings to previously studied heterostructures made from exfoliated flakes\cite{Dufferwiel2015,Sidler2017,Dufferwiel2017,Dufferwiel2018}.

Furthermore, the demonstrated CVD growth on sapphire and hBN produced highly orientated TMD islands, and is thus suitable for the fabrication of large scale TMD/TMD heterostructures with highly controlled interlayer twist angle\cite{Alexeev2019} to be embedded in microcavities. Together with additional hBN and graphene layers these structures could provide a viable route to realization of highly tunable and non-linear dipolar polaritons \cite{Cristofolini2012} in large scale devices.  

This work demonstrates the possibility to fabricate large scale polaritonic devices based on van-der-Waals heterostructures. Further development of large scale monolayer semiconductor growth techniques, most notably directly onto hBN which provides highly co-orientated TMD domains, will inevitably lead to heterostructures that can reliably and repeatedly compete with, or out-perform, those built with exfoliated flakes due to the unprecedented scalability that is granted.

\section*{Methods}

{\bf Dielectric mirror fabrication.}
Highly reflecting distributed Bragg reflectors (DBRs) are deposited on silica substrates by ion beam sputtering. The DBRs are comprised of 13 pairs of quarter wavelength SiO$_2$/Ta$_2$O$_5$ layers of thicknesses 129 and 89 nm (refractive index 1.45 and 2.10 respectively), terminating with SiO$_2$. The DBRs are designed for a center wavelength of 750 nm and a stop-band width of 200 nm.

The concave-shaped template for the top mirror is produced by focused ion beam milling in a smooth fused silica substrate. Gallium ions are accelerated onto a precise position of the silica substrate achieving an accuracy of around 5 nm with an r.m.s. roughness below 1 nm \cite{Dolan2010}. The nominal radius of curvature of the concave mirror was 20 $\mu$m, leading to a beam waist on the planar mirror of around 1 $\mu$m \cite{Schwarz2014,Dufferwiel2015}.

\medskip
{\bf Growth of single layer MoSe$_2$ and transfer to SiO$_2$/Si.}
MoSe$_2$ was grown on c-plane sapphire by CVD. Two precursors, MoO$_3$ (99.97\%, Sigma Aldrich) and Se (99.999\%, Alfa Aesar), were used for the growth. 150 mg of Se was placed at the upstream entry of the furnace and 60 mg of MoO$_3$ powder was placed at the centre of the furnace. A crucible containing MoO$_3$ was partially covered by a SiO$_2$/Si wafer to reduce intense evaporation of the precursor. The sapphire substrate was located next to the crucible that contained MoO$_3$. Before the tube furnace was heated, the tube was evacuated for 30 min and filled with the Ar gas achieving ambient pressure. The temperature of the furnace was increased up to 600 $\degree$C for 18 min under a steady flow of Ar gas (60 sccm) and H$_2$ gas (12 sccm). When the furnace reached 600 $\degree$C, Se was vaporized by heating the upstream entry of the tube up to 270 $\degree$C using a heating belt. Finally, temperature of the tube furnace was increased to 700 $\degree$C and maintained for 1 hr for the MoSe$_2$ growth. Afterwards, the tube furnace was cooled down to room temperature while the Ar flow was maintained without H$_2$. To transfer MoSe$_2$ on top of the DBR, polystyrene was used to maintain the sample quality.

\medskip
{\bf Growth of single layer WSe$_2$ on hBN.} 
To fabricate WSe$_2$ on hBN, multilayer hBN was initially grown on c-plane sapphire (see Methods). WO$_3$ (99.998\%, Alfa Aesar) and Se (99.999\%, Alfa Aesar), were used for the WSe$_2$ growth. 300 mg  of Se was placed at the upstream entry of the furnace and 120 mg of WO$_3$ powder was placed at the center of the furnace. To reduce the influence of humidity, a small amount of NaCl was mixed with WO$_3$ powder. The multilayer hBN on sapphire substrate was positioned next to the crucible containing WO$_3$. Before the tube furnace was heated, the tube was evacuated for more than 30 min. Then, the temperature of the tube furnace was increased to 800 $\degree$C for 24 min under a steady flow of Ar gas (120 sccm) and H$_2$ gas (20 sccm). When the furnace reached 800 $\degree$C, the Se was vaporized by heating the upstream entry of the tube to 270 $\degree$C using a heating belt. Finally, temperature of the tube furnace was increased to 870 $\degree$C and maintained for 1 hour for the WSe$_2$ growth. Afterwards, the tube furnace was cooled down to room temperature under Ar flow.

\medskip
{\bf Growth of large area hBN.} 
Multilayer hBN with an AA' stacking order was grown by remote inductively coupled plasma chemical vapor deposition method. A 2-inch c-plane sapphire was used as a substrate for the hBN growth. The substrate was placed in the center of a 2-inch alumina tube furnace of CVD. A borazine (Gelest, Inc.) precursor flask was placed in a water bath at -15 $\degree$C. The bath temperature before the growth of hBN was increased up to 25 $\degree$C. Before the growth of multilayer hBN, the furnace was heated to 1220 $\degree$C under flow of Ar gas (10 sccm). Plasma was generated at a power of 30 W under a flow of borazine (0.2 sccm) and Ar (10 sccm) gases for 30 mins. Atomic force microscopy and transmission electron microscopy measurements confirmed that the thickness of hBN was 1.2 nm, approximately 3 layers. In addition, the hBN sample shows quite good thickness uniformity over the 2-inch sapphire substrate according to the Raman and UV absorption spectra measured at nine random points over the 2-inch hBN film.

\medskip
{\bf Optical measurements.}
The photoluminescence images of the CVD samples were acquired using a modified bright-field microscope (LV150N, Nikon) equipped with a color camera (DS-Vi1, Nikon). The near-infrared emission from the white-light source was blocked with a 550-nm short-pass filter (FESH0550, Thorlabs), and a 600-nm long-pass filter (FELH0600, Thorlabs) was used to isolate the photoluminescence signal from the samples. The full description of the system is available in Ref.\cite{Alexeev2017}.

Spectrally-resolved photoluminescence (PL) and reflectance contrast (RC) measurements were performed using a custom-built micro-PL setup. For PL, the excitation light centered at 2.33 eV and 1.88 eV was generated by two diode-pumped solid-state lasers (CW532-050 and ADL-66505TL, Roithner). For RC, a stabilized tungsten-halogen white-light source (SLS201L, Thorlabs) was used. The excitation light was focused onto the sample using a 50x objective lens (M Plan Apo 50X, Mitutoyo). The PL and RC signals collected in the backwards direction were detected by a 0.5 m spectrometer (SP-2-500i, Princeton Instruments) with a nitrogen-cooled charge-coupled device camera (PyLoN:100BR, Princeton Instruments). The PL signal was isolated using 700 nm and 650 nm long-pass filters (FEL0700 and FEL0650, Thorlabs). The RC spectra were derived by comparing the spectra of white light reflected from the sample and the substrate as $\Delta$ $R/R= (R_{sub}-R_{HS}) / R_{sub}$, where $R_{HS}$ ($R_{sub}$) is the intensity of light reflected by the sample (substrate). The room-temperature measurements were performed in ambient conditions. The low-temperature measurements were carried out using a continuous-flow liquid helium cryostat, in which the sample was placed on a cold finger with a base temperature of $\sim$ 5 K.

\medskip
{\bf Tunable micro-cavity.}
The optical cavity was formed using an external concave mirror, with nominal radius of 20 $\mu$m, to produce a 0D tunable cavity \cite{Schwarz2014}.

The bottom mirror is controlled by a 5-axis piezo-actuator stack, the first three stages control the {\it x}, {\it y}, and {\it z} translational motion, while another two stages control the tilt alignment. The top mirror is positioned using a 3-axis piezo-actuator stage controlling the {\it x}, {\it y}, and {\it z} translational motion. Optical PL scans were completed with the samples placed in a helium bath cryostat system at a temperature of 4.2 K using a 637 nm continuous-wave laser diode (Vortran Stradus), focused onto the sample using an achromatic lens. The collected PL is focused onto a single mode fiber and guided into a 0.75 m spectrometer (SP-2-750i, Princeton Instruments) and a high-sensitivity charge-coupled device (PyLoN:400BR, Princeton Instruments) for emission collection.

\medskip
{\bf Measurement of monolithic cavities.}
We performed the Fourier space spectral imaging of the PL emitted by the monolithic cavity by employing a 2D CCD array (PyLoN:100BR, Princeton Instruments) coupled to a 300 gr/mm grating spectrometer (SP-2-500i, Princeton Instruments). We focused a 30 cm lens onto the back plane of a 50x Mitutoyo infinity corrected objective to obtain the Fourier plane image of the sample, which was then projected on the slit of the spectrometer by using a 10 cm lens. We used the slit to select only the section of the Fourier space at k$_x$ = 0, resulting in a final image on the CCD displaying the PL as a function of k$_y$ on the y-axis and energy on the x-axis. The conversion from k$_y$ to angles has been carried out by considering k$_y \approx \sin\theta$ and knowing that the maximum k$_y$ detected by our setup is equal to the objective NA = 0.55

\medskip
{\bf Flake orientation analysis.}
Optical microscope PL images were analyzed in MATLAB using functions from the image processing toolbox \cite{matlabImage}. The color thresholding application was used to isolate monolayer material in a typical PL image and was applied to each combination of monolayer and substrate. The program identified 8089 triangular objects in HS1 and 8391 triangular objects in HS2. Further details of the image processing and a more complete explanation of the analysis can be found in Ref.\cite{SeversMillard2020}.

\medskip
{\bf Fabrication of monolithic cavities.}
The monolithic cavity has been fabricated by depositing a SiO$_2$ film of 98nm on top of the CVD-grown monolayers, which were previously transferred on a 13 pairs SiO$_2$/Ta$_2$O$_5$ DBR grown by ion beam assisted sputtering on a sapphire substrate. In order to minimize the potential damages on the monolayers, the silica layer covering the TMDs has been grown at room temperature by using an e-beam deposition system. For the top mirror, a semi-transparent layer of Au (thickness: 50 nm) has been thermally evaporated, completing the cavity.


\section*{Acknowledgments}
D. J. G.,  A.G., T.S.M., and A.I.T. acknowledge funding by EPSRC (EP/P026850/1).
This work was supported by the research funds (NRF-2019R1A4A1027934 and NRF-2017R1E1A1A01074493) through National Research Foundation by the Ministry of Science and ICT, Korea.
R. J., K. G., and D. G. L. thank the financial support of EPSRC via programme grant 'Hybrid Polaritonics' (EP/M025330/1).
T. P. L. acknowledges financial support from the EPSRC Doctoral Prize Fellowship scheme Grant Reference EP/R513313/1. A.I.T. thanks the financial support of the Graphene Flagship under grant agreements 785219.

\subsection*{Author Contributions}

D. J. G., A. G., and S. A. contributed equally to this work. D. J. G., A. G., and T. P. L. carried out optical investigations. TMD monolayers were grown via CVD, and samples fabricated, by S. A. and H. S. S. hBN layers were grown via CVD by K. Y. M. and A-R. J. The monolithic cavity was completed by R. J., K. G., and D. G. L. The concave mirrors were made by A. A. P. T. and J. M. S. Data was analyzed by D. J. G. and A. G. The manuscript was written by D. J. G. with major input from A. G. and further contributions from all co-authors. A. I. T. provided management of various aspects of the project, and contributed to the analysis and interpretation of the data and writing of the manuscript. A. I. T., H. S. S., D. G. L., and J. M. S. provided management of relevant parts of the project. A. I. T. conceived and oversaw the whole project.

\subsection*{Data Availability}
\noindent
The data that support the plots within this paper and other findings of this study are available from the corresponding author upon reasonable request.


\subsection*{Competing interests}
\noindent
The authors declare no competing interests.

\subsection*{Supplementary Information}
\noindent
Supplementary information are available upon reasonable request.

\def\bibsection{\section*{Bibliography}}
\setstretch{1.2}
\bibliography{CVDPaperBib_LSA.bib}

\begin{thebibliography}{10}
\expandafter\ifx\csname url\endcsname\relax
  \def\url#1{\texttt{#1}}\fi
\expandafter\ifx\csname urlprefix\endcsname\relax\def\urlprefix{URL }\fi
\providecommand{\bibinfo}[2]{#2}
\providecommand{\eprint}[2][]{\url{#2}}

\bibitem{Novoselov2005}
\bibinfo{author}{Novoselov, K.~S.} \emph{et~al.}
\newblock \bibinfo{title}{{Two-dimensional atomic crystals}}.
\newblock \emph{\bibinfo{journal}{Proceedings of the National Academy of
  Sciences of the United States of America}} \textbf{\bibinfo{volume}{102}},
  \bibinfo{pages}{10451--10453} (\bibinfo{year}{2005}).

\bibitem{Mak2016}
\bibinfo{author}{Mak, K.~F.} \& \bibinfo{author}{Shan, J.}
\newblock \bibinfo{title}{{Photonics and optoelectronics of 2D semiconductor
  transition metal dichalcogenides}}.
\newblock \emph{\bibinfo{journal}{Nature Photonics}}
  \textbf{\bibinfo{volume}{10}}, \bibinfo{pages}{216--226}
  (\bibinfo{year}{2016}).

\bibitem{Mak2013}
\bibinfo{author}{Mak, K.~F.} \emph{et~al.}
\newblock \bibinfo{title}{{Tightly bound trions in monolayer MoS 2}}.
\newblock \emph{\bibinfo{journal}{Nature Materials}}
  \textbf{\bibinfo{volume}{12}}, \bibinfo{pages}{207--211}
  (\bibinfo{year}{2013}).

\bibitem{He2014}
\bibinfo{author}{He, K.} \emph{et~al.}
\newblock \bibinfo{title}{{Tightly bound excitons in monolayer WSe2}}.
\newblock \emph{\bibinfo{journal}{Physical Review Letters}}
  \textbf{\bibinfo{volume}{113}}, \bibinfo{pages}{1--5} (\bibinfo{year}{2014}).

\bibitem{Xu2014}
\bibinfo{author}{Xu, X.}, \bibinfo{author}{Yao, W.}, \bibinfo{author}{Xiao, D.}
  \& \bibinfo{author}{Heinz, T.~F.}
\newblock \bibinfo{title}{{Spin and pseudospins in layered transition metal
  dichalcogenides}}.
\newblock \emph{\bibinfo{journal}{Nature Physics}}
  \textbf{\bibinfo{volume}{10}}, \bibinfo{pages}{343--350}
  (\bibinfo{year}{2014}).

\bibitem{Liu2015}
\bibinfo{author}{Liu, X.} \emph{et~al.}
\newblock \bibinfo{title}{{Strong light–matter coupling in two-dimensional
  atomic crystals}}.
\newblock \emph{\bibinfo{journal}{Nature Photonics}}
  \textbf{\bibinfo{volume}{9}}, \bibinfo{pages}{30--34} (\bibinfo{year}{2014}).

\bibitem{Dhara2018}
\bibinfo{author}{Dhara, S.} \emph{et~al.}
\newblock \bibinfo{title}{{Anomalous dispersion of microcavity
  trion-polaritons}}.
\newblock \emph{\bibinfo{journal}{Nature Physics}}
  \textbf{\bibinfo{volume}{14}}, \bibinfo{pages}{130--133}
  (\bibinfo{year}{2018}).

\bibitem{Dufferwiel2015}
\bibinfo{author}{Dufferwiel, S.} \emph{et~al.}
\newblock \bibinfo{title}{{Exciton-polaritons in van der Waals heterostructures
  embedded in tunable microcavities}}.
\newblock \emph{\bibinfo{journal}{Nature Communications}}
  \textbf{\bibinfo{volume}{6}}, \bibinfo{pages}{1--7} (\bibinfo{year}{2015}).

\bibitem{Sidler2017}
\bibinfo{author}{Sidler, M.} \emph{et~al.}
\newblock \bibinfo{title}{{Fermi polaron-polaritons in charge-tunable
  atomically thin semiconductors}}.
\newblock \emph{\bibinfo{journal}{Nature Physics}}
  \textbf{\bibinfo{volume}{13}}, \bibinfo{pages}{255--261}
  (\bibinfo{year}{2017}).

\bibitem{Low2016}
\bibinfo{author}{Low, T.} \emph{et~al.}
\newblock \bibinfo{title}{{Polaritons in layered two-dimensional materials}}.
\newblock \emph{\bibinfo{journal}{Nature Materials}}
  \textbf{\bibinfo{volume}{16}}, \bibinfo{pages}{182--194}
  (\bibinfo{year}{2017}).

\bibitem{Schneider2018}
\bibinfo{author}{Schneider, C.} \emph{et~al.}
\newblock \bibinfo{title}{{Two-dimensional semiconductors in the regime of
  strong light-matter coupling}}.
\newblock \emph{\bibinfo{journal}{Nature Communications}}
  \textbf{\bibinfo{volume}{9}}, \bibinfo{pages}{2695} (\bibinfo{year}{2018}).

\bibitem{Krol2020}
\bibinfo{author}{Krol, M.} \emph{et~al.}
\newblock \bibinfo{title}{{Exciton-polaritons in multilayer WSe 2 in a planar
  microcavity}}.
\newblock \emph{\bibinfo{journal}{2D Materials}} \textbf{\bibinfo{volume}{7}},
  \bibinfo{pages}{15006} (\bibinfo{year}{2020}).

\bibitem{Dufferwiel2017}
\bibinfo{author}{Dufferwiel, S.} \emph{et~al.}
\newblock \bibinfo{title}{{Valley-addressable polaritons in atomically thin
  semiconductors}}.
\newblock \emph{\bibinfo{journal}{Nature Photonics}}
  \textbf{\bibinfo{volume}{11}}, \bibinfo{pages}{497--501}
  (\bibinfo{year}{2017}).

\bibitem{Chen2017}
\bibinfo{author}{Chen, Y.-J.}, \bibinfo{author}{Cain, J.~D.},
  \bibinfo{author}{Stanev, T.~K.}, \bibinfo{author}{Dravid, V.~P.} \&
  \bibinfo{author}{Stern, N.~P.}
\newblock \bibinfo{title}{{Valley-polarized exciton–polaritons in a monolayer
  semiconductor}}.
\newblock \emph{\bibinfo{journal}{Nature Photonics}}
  \textbf{\bibinfo{volume}{11}}, \bibinfo{pages}{431--435}
  (\bibinfo{year}{2017}).

\bibitem{Dufferwiel2018}
\bibinfo{author}{Dufferwiel, S.} \emph{et~al.}
\newblock \bibinfo{title}{{Valley coherent exciton-polaritons in a monolayer
  semiconductor}}.
\newblock \emph{\bibinfo{journal}{Nature communications}}
  \textbf{\bibinfo{volume}{9}}, \bibinfo{pages}{4797} (\bibinfo{year}{2018}).

\bibitem{Qiu2019}
\bibinfo{author}{Qiu, L.}, \bibinfo{author}{Chakraborty, C.},
  \bibinfo{author}{Dhara, S.} \& \bibinfo{author}{Vamivakas, A.~N.}
\newblock \bibinfo{title}{{Room-temperature valley coherence in a polaritonic
  system}}.
\newblock \emph{\bibinfo{journal}{Nature Communications}}
  \textbf{\bibinfo{volume}{10}}, \bibinfo{pages}{1513} (\bibinfo{year}{2019}).

\bibitem{barachati2018interacting}
\bibinfo{author}{Barachati, F.} \emph{et~al.}
\newblock \bibinfo{title}{{Interacting polariton fluids in a monolayer of
  tungsten disulfide}}.
\newblock \emph{\bibinfo{journal}{Nature nanotechnology}}
  \textbf{\bibinfo{volume}{13}}, \bibinfo{pages}{906--909}
  (\bibinfo{year}{2018}).

\bibitem{lundt2019optical}
\bibinfo{author}{Lundt, N.} \emph{et~al.}
\newblock \bibinfo{title}{{Optical valley Hall effect for highly
  valley-coherent exciton-polaritons in an atomically thin semiconductor}}.
\newblock \emph{\bibinfo{journal}{Nature nanotechnology}}
  \textbf{\bibinfo{volume}{14}}, \bibinfo{pages}{770--775}
  (\bibinfo{year}{2019}).

\bibitem{Waldherr2018}
\bibinfo{author}{Waldherr, M.} \emph{et~al.}
\newblock \bibinfo{title}{{Observation of bosonic condensation in a hybrid
  monolayer MoSe2-GaAs microcavity}}.
\newblock \emph{\bibinfo{journal}{Nature Communications}}
  \textbf{\bibinfo{volume}{9}}, \bibinfo{pages}{3286} (\bibinfo{year}{2018}).

\bibitem{Emmanuele2020}
\bibinfo{author}{Emmanuele, R. P.~A.} \emph{et~al.}
\newblock \bibinfo{title}{{Highly nonlinear trion-polaritons in a monolayer
  semiconductor}}.
\newblock \emph{\bibinfo{journal}{Nature Communications}}
  \textbf{\bibinfo{volume}{11}}, \bibinfo{pages}{3589} (\bibinfo{year}{2020}).

\bibitem{Kasprzak2006}
\bibinfo{author}{Kasprzak, J.} \emph{et~al.}
\newblock \bibinfo{title}{{Bose-Einstein condensation of exciton polaritons}}.
\newblock \emph{\bibinfo{journal}{Nature}} \textbf{\bibinfo{volume}{443}},
  \bibinfo{pages}{409--414} (\bibinfo{year}{2006}).

\bibitem{Christopoulos2007}
\bibinfo{author}{Christopoulos, S.} \emph{et~al.}
\newblock \bibinfo{title}{{Room-temperature polariton lasing in semiconductor
  microcavities}}.
\newblock \emph{\bibinfo{journal}{Physical Review Letters}}
  \textbf{\bibinfo{volume}{98}}, \bibinfo{pages}{126405}
  (\bibinfo{year}{2007}).

\bibitem{Bhattacharya2014}
\bibinfo{author}{Bhattacharya, P.} \emph{et~al.}
\newblock \bibinfo{title}{{Room temperature electrically injected polariton
  laser}}.
\newblock \emph{\bibinfo{journal}{Physical Review Letters}}
  \textbf{\bibinfo{volume}{112}}, \bibinfo{pages}{236802}
  (\bibinfo{year}{2014}).

\bibitem{Amo2009}
\bibinfo{author}{Amo, A.} \emph{et~al.}
\newblock \bibinfo{title}{{Superfluidity of polaritons in semiconductor
  microcavities}}.
\newblock \emph{\bibinfo{journal}{Nature Physics}}
  \textbf{\bibinfo{volume}{5}}, \bibinfo{pages}{805--810}
  (\bibinfo{year}{2009}).

\bibitem{Liew2010}
\bibinfo{author}{Liew, T. C.~H.} \emph{et~al.}
\newblock \bibinfo{title}{{Exciton-polariton integrated circuits}}.
\newblock \emph{\bibinfo{journal}{Physical Review B}}
  \textbf{\bibinfo{volume}{82}}, \bibinfo{pages}{33302} (\bibinfo{year}{2010}).

\bibitem{Ballarini2013}
\bibinfo{author}{Ballarini, D.} \emph{et~al.}
\newblock \bibinfo{title}{{All-optical polariton transistor}}.
\newblock \emph{\bibinfo{journal}{Nature Communications}}
  \textbf{\bibinfo{volume}{4}} (\bibinfo{year}{2013}).

\bibitem{Shree2019}
\bibinfo{author}{Shree, S.} \emph{et~al.}
\newblock \bibinfo{title}{{High optical quality of MoS 2 monolayers grown by
  chemical vapor deposition}}.
\newblock \emph{\bibinfo{journal}{2D Materials}} \textbf{\bibinfo{volume}{7}},
  \bibinfo{pages}{15011} (\bibinfo{year}{2019}).

\bibitem{Zhang2019a}
\bibinfo{author}{Zhang, Y.} \emph{et~al.}
\newblock \bibinfo{title}{{Recent Progress in CVD Growth of 2D Transition Metal
  Dichalcogenides and Related Heterostructures}}.
\newblock \emph{\bibinfo{journal}{Advanced Materials}}
  \textbf{\bibinfo{volume}{31}}, \bibinfo{pages}{1901694}
  (\bibinfo{year}{2019}).

\bibitem{Gebhardt2018}
\bibinfo{author}{Gebhardt, C.} \emph{et~al.}
\newblock \bibinfo{title}{{Polariton hyperspectral imaging of two-dimensional
  semiconductor crystals}}.
\newblock \emph{\bibinfo{journal}{Scientific Reports}}
  \textbf{\bibinfo{volume}{9}}, \bibinfo{pages}{13756} (\bibinfo{year}{2019}).

\bibitem{SeversMillard2020}
\bibinfo{author}{{Severs Millard}, T.} \emph{et~al.}
\newblock \bibinfo{title}{{Large area chemical vapour deposition grown
  transition metal dichalcogenide monolayers automatically characterized
  through photoluminescence imaging}}.
\newblock \emph{\bibinfo{journal}{npj 2D Materials and Applications}}
  \textbf{\bibinfo{volume}{4}}, \bibinfo{pages}{12} (\bibinfo{year}{2020}).

\bibitem{Ajayi2017}
\bibinfo{author}{Ajayi, O.~A.} \emph{et~al.}
\newblock \bibinfo{title}{{Approaching the intrinsic photoluminescence
  linewidth in transition metal dichalcogenide monolayers}}.
\newblock \emph{\bibinfo{journal}{2D Materials}} \textbf{\bibinfo{volume}{4}},
  \bibinfo{pages}{31011} (\bibinfo{year}{2017}).

\bibitem{DelPozoZamudio2020}
\bibinfo{author}{{Del Pozo-Zamudio}, O.} \emph{et~al.}
\newblock \bibinfo{title}{{Electrically pumped WSe 2 -based light-emitting van
  der Waals heterostructures embedded in monolithic dielectric microcavities}}.
\newblock \emph{\bibinfo{journal}{2D Materials}} \textbf{\bibinfo{volume}{7}},
  \bibinfo{pages}{31006} (\bibinfo{year}{2020}).

\bibitem{Zhang2019}
\bibinfo{author}{Zhang, X.} \emph{et~al.}
\newblock \bibinfo{title}{{Defect-Controlled Nucleation and Orientation of WSe
  2 on hBN: A Route to Single-Crystal Epitaxial Monolayers}}.
\newblock \emph{\bibinfo{journal}{ACS Nano}} \textbf{\bibinfo{volume}{13}},
  \bibinfo{pages}{3341--3352} (\bibinfo{year}{2019}).

\bibitem{Okada2014}
\bibinfo{author}{Okada, M.} \emph{et~al.}
\newblock \bibinfo{title}{{Direct chemical vapor deposition growth of WS2
  atomic layers on hexagonal boron nitride}}.
\newblock \emph{\bibinfo{journal}{ACS Nano}} \textbf{\bibinfo{volume}{8}},
  \bibinfo{pages}{8273--8277} (\bibinfo{year}{2014}).

\bibitem{Yu2017a}
\bibinfo{author}{Yu, H.} \emph{et~al.}
\newblock \bibinfo{title}{{Precisely Aligned Monolayer MoS 2 Epitaxially Grown
  on h-BN basal Plane}}.
\newblock \emph{\bibinfo{journal}{Small}} \textbf{\bibinfo{volume}{13}},
  \bibinfo{pages}{1603005} (\bibinfo{year}{2017}).

\bibitem{Ross2013}
\bibinfo{author}{Ross, J.~S.} \emph{et~al.}
\newblock \bibinfo{title}{{Electrical control of neutral and charged excitons
  in a monolayer semiconductor}}.
\newblock \emph{\bibinfo{journal}{Nature Communications}}
  \textbf{\bibinfo{volume}{4}}, \bibinfo{pages}{1473--1476}
  (\bibinfo{year}{2013}).

\bibitem{Robert2017}
\bibinfo{author}{Robert, C.} \emph{et~al.}
\newblock \bibinfo{title}{{Fine structure and lifetime of dark excitons in
  transition metal dichalcogenide monolayers}}.
\newblock \emph{\bibinfo{journal}{Physical Review B}}
  \textbf{\bibinfo{volume}{96}}, \bibinfo{pages}{1--8} (\bibinfo{year}{2017}).

\bibitem{Lindlau2018}
\bibinfo{author}{Lindlau, J.} \emph{et~al.}
\newblock \bibinfo{title}{{The role of momentum-dark excitons in the elementary
  optical response of bilayer WSe2}}.
\newblock \emph{\bibinfo{journal}{Nature Communications}}
  \textbf{\bibinfo{volume}{9}}, \bibinfo{pages}{2586} (\bibinfo{year}{2018}).

\bibitem{Jadczak2017}
\bibinfo{author}{Jadczak, J.} \emph{et~al.}
\newblock \bibinfo{title}{{Probing of free and localized excitons and trions in
  atomically thin WSe 2 , WS 2 , MoSe 2 and MoS 2 in photoluminescence and
  reflectivity experiments}}.
\newblock \emph{\bibinfo{journal}{Nanotechnology}}
  \textbf{\bibinfo{volume}{28}}, \bibinfo{pages}{395702}
  (\bibinfo{year}{2017}).

\bibitem{Lippert2017}
\bibinfo{author}{Lippert, S.} \emph{et~al.}
\newblock \bibinfo{title}{{Influence of the substrate material on the optical
  properties of tungsten diselenide monolayers}}.
\newblock \emph{\bibinfo{journal}{2D Materials}} \textbf{\bibinfo{volume}{4}},
  \bibinfo{pages}{025045} (\bibinfo{year}{2017}).

\bibitem{Dumcenco2015}
\bibinfo{author}{Dumcenco, D.} \emph{et~al.}
\newblock \bibinfo{title}{{Large-Area Epitaxial Monolayer MoS 2}}.
\newblock \emph{\bibinfo{journal}{ACS Nano}} \textbf{\bibinfo{volume}{9}},
  \bibinfo{pages}{4611--4620} (\bibinfo{year}{2015}).

\bibitem{Zhang2018}
\bibinfo{author}{Zhang, X.} \emph{et~al.}
\newblock \bibinfo{title}{{Diffusion-Controlled Epitaxy of Large Area Coalesced
  WSe 2 Monolayers on Sapphire}}.
\newblock \emph{\bibinfo{journal}{Nano Letters}} \textbf{\bibinfo{volume}{18}},
  \bibinfo{pages}{1049--1056} (\bibinfo{year}{2018}).

\bibitem{Schwarz2014}
\bibinfo{author}{Schwarz, S.} \emph{et~al.}
\newblock \bibinfo{title}{{Two-dimensional metal-chalcogenide films in tunable
  optical microcavities}}.
\newblock \emph{\bibinfo{journal}{Nano Letters}} \textbf{\bibinfo{volume}{14}},
  \bibinfo{pages}{7003--7008} (\bibinfo{year}{2014}).

\bibitem{Kavokin2008}
\bibinfo{author}{Kavokin, A.}, \bibinfo{author}{Baumberg, J.~J.},
  \bibinfo{author}{Malpuech, G.} \& \bibinfo{author}{Laussy, F.~P.}
\newblock \emph{\bibinfo{title}{{Microcavities}}} (\bibinfo{publisher}{Oxford
  University Press}, \bibinfo{year}{2007}).

\bibitem{GENCO2018174}
\bibinfo{author}{Genco, A.} \emph{et~al.}
\newblock \bibinfo{title}{{High quality factor microcavity OLED employing
  metal-free electrically active Bragg mirrors}}.
\newblock \emph{\bibinfo{journal}{Organic Electronics}}
  \textbf{\bibinfo{volume}{62}}, \bibinfo{pages}{174--180}
  (\bibinfo{year}{2018}).

\bibitem{Alexeev2019}
\bibinfo{author}{Alexeev, E.~M.} \emph{et~al.}
\newblock \bibinfo{title}{{Resonantly hybridized excitons in moir{\'{e}}
  superlattices in van der Waals heterostructures}}.
\newblock \emph{\bibinfo{journal}{Nature}} \textbf{\bibinfo{volume}{567}},
  \bibinfo{pages}{81--86} (\bibinfo{year}{2019}).

\bibitem{Cristofolini2012}
\bibinfo{author}{Cristofolini, P.} \emph{et~al.}
\newblock \bibinfo{title}{{Coupling Quantum Tunneling with Cavity Photons}}.
\newblock \emph{\bibinfo{journal}{Science}} \textbf{\bibinfo{volume}{336}},
  \bibinfo{pages}{704--707} (\bibinfo{year}{2012}).

\bibitem{Dolan2010}
\bibinfo{author}{Dolan, P.~R.}, \bibinfo{author}{Hughes, G.~M.},
  \bibinfo{author}{Grazioso, F.}, \bibinfo{author}{Patton, B.~R.} \&
  \bibinfo{author}{Smith, J.~M.}
\newblock \bibinfo{title}{{Femtoliter tunable optical cavity arrays}}.
\newblock \emph{\bibinfo{journal}{Optics Letters}}
  \textbf{\bibinfo{volume}{35}}, \bibinfo{pages}{3556} (\bibinfo{year}{2010}).

\bibitem{Alexeev2017}
\bibinfo{author}{Alexeev, E.~M.} \emph{et~al.}
\newblock \bibinfo{title}{{Imaging of Interlayer Coupling in van der Waals
  Heterostructures Using a Bright-Field Optical Microscope}}.
\newblock \emph{\bibinfo{journal}{Nano Letters}} \textbf{\bibinfo{volume}{17}},
  \bibinfo{pages}{5342--5349} (\bibinfo{year}{2017}).

\bibitem{matlabImage}
\bibinfo{author}{MathWorks}.
\newblock \emph{\bibinfo{title}{{Image Processing Toolbox User's Guide}}}
  (\bibinfo{year}{2017}).

\end{thebibliography}
\end{document}